\documentclass[smallextended]{svjour3}     

\smartqed  

\usepackage[hyphens]{url}  
\usepackage{breakurl}  

\usepackage{pgfplots}
\pgfplotsset{compat=1.17}

\usepackage{lineno}
\usepackage[colorlinks=true]{hyperref}
\modulolinenumbers[5]

\usepackage{booktabs}   
\usepackage{subcaption} 

\usepackage[labelfont=bf]{caption}
\usepackage[utf8]{inputenc} 
\usepackage[T1]{fontenc}
\usepackage{microtype}

\usepackage{enumitem}
\usepackage{color}
\usepackage{amssymb}
\usepackage{multirow}
\usepackage{etoolbox}
\usepackage[plain]{algorithm}
\usepackage[noend]{algorithmic}
\usepackage{syntax}
\usepackage{comment}
\usepackage{amsmath}
\usepackage{xspace}
\usepackage{tcolorbox}
\usepackage[all]{nowidow}
\usepackage{booktabs}
\usepackage{algorithmic}
\usepackage[english]{babel}
\usepackage{hyphenat}
\usepackage{ulem}
\usepackage{listings}
\usepackage{mdframed}

\definecolor{pblue}{rgb}{0.13,0.13,1}
\definecolor{pgreen}{rgb}{0,0.5,0}
\definecolor{pred}{rgb}{0.9,0,0}
\definecolor{pgrey}{rgb}{0.46,0.45,0.48}
\definecolor{darkblue}{rgb}{0.0, 0.0, 0.55}
\definecolor{light-gray}{gray}{0.9}

\newcommand{\lstbg}[3][0pt]{{\fboxsep#1\colorbox{#2}{\strut #3}}}

\lstdefinelanguage{diff}{
  basicstyle=\ttfamily\small,
  morecomment=[f][\lstbg{pred!20}]-,
  morecomment=[f][\lstbg{pgreen!20}]+,
  morecomment=[f][\textit]{@@},
  morecomment=[f][\textit]{---},
  morecomment=[f][\textit]{+++},
}

\hypersetup{
    colorlinks=true,
    linkcolor=blue,
    filecolor=magenta,      
    urlcolor=blue,
    citecolor=violet,
    pdfpagemode=FullScreen,
}

\usepackage{listings}

\newcommand{\bugs}{100}
\newcommand{\bugsCE}{51}
\newcommand{\bugsBC}{19}

\newcommand{\refactoringTypesOWC}{15}
\newcommand{\refactoringTypesOSC}{10}

\newcommand{\slms}{\llama{}, \mistral{}, \gemma{}, \gemmaNew{}, \deepseek{}, \phimodel{}, \omini{}, and \ominihigh{}}
\newcommand{\gemma}{\textsc{Gemma 2 9B}}
\newcommand{\gemmaNew}{\textsc{Gemma 3 12B}}
\newcommand{\llama}{\textsc{Llama 3.2 3B}}
\newcommand{\phimodel}{\textsc{Phi-4 14B}}
\newcommand{\deepseek}{\textsc{DeepSeek-R1 14B}}
\newcommand{\omini}{\textsc{o1-mini}}
\newcommand{\ominihigh}{\textsc{o3-mini-high}}
\newcommand{\mistral}{\textsc{Mistral 7B}}
\newcommand{\gpt}{\textsc{ChatGPT}}
\newcommand{\claude}{\textsc{Claude}}

\definecolor{green}{RGB}{0,100,0}

\newcommand{\eclipse}{\textsc{Eclipse}}
\newcommand{\jrrt}{\textsc{JRRT}}
\newcommand{\netbeans}{\textsc{NetBeans}}
\newcommand{\intellij}{\textsc{IntelliJ}}
\newcommand{\vscode}{\textsc{VSCode}}
\newcommand{\rope}{\textsc{Rope}}

\newcommand{\saferefactor}{\textsc{SafeRefactor}}
\newcommand{\jdolly}{\textsc{JDolly}}

\journalname{}

\begin{document}

\title{
Evaluating the Effectiveness of Small Language Models in Detecting Refactoring Bugs
}

\author{Rohit Gheyi \and M\'{a}rcio Ribeiro \and Jonhnanthan Oliveira
}

\institute{R. Gheyi and J. Oliveira \at
              Federal University of Campina Grande \\
              \email{ rohit@dsc.ufcg.edu.br, jonhnanthan@copin.ufcg.edu.br}             \\
           \and
           M. Ribeiro\at
           Federal University of Alagoas \\
           \email{marcio@ic.ufal.br}
}  

\date{}

\maketitle

\begin{abstract}

Popular Integrated Development Environments (IDEs) frequently contain bugs in their refactoring implementations. Ensuring that a transformation preserves a program’s behavior is a complex task. Traditional detection methods rely on predefined preconditions for each refactoring type, limiting their scalability and adaptability to new transformations. These methods often require extensive static and dynamic analyses, which are computationally expensive, time-consuming, and may still fail to detect certain refactoring bugs.
This study evaluates the effectiveness of Small Language Models (SLMs) in detecting two types of refactoring bugs in Java and Python: (i) transformations that introduce errors or behavioral changes (\textsc{Type I}) and (ii) transformations unnecessarily blocked by IDEs despite being valid (\textsc{Type II}).
We assess whether \slms{} can accurately detect \bugs{} refactoring bugs reported in widely used Java and Python IDEs, such as \eclipse{} and \netbeans{}. The study covers 16 refactoring types and employs zero-shot prompting on consumer-grade hardware to evaluate the models’ ability to reason about refactoring correctness without explicit prior training.
The proprietary \ominihigh{} model achieved the highest detection rate, identifying 84.3\% of \textsc{Type I} bugs. The open-source \phimodel{} performed comparably well, demonstrating strong effectiveness across both bug types. However, \ominihigh{} struggled with \textsc{Type II} bugs, correctly identifying and applying valid but blocked transformations in only 40\% of cases. Some open models, such as \phimodel{}, exhibited comparable performance, underscoring a broader challenge in effectively addressing these issues.
The findings highlight the potential of SLMs for efficiently detecting refactoring bugs, particularly in verifying behavioral changes. Integrating SLMs into refactoring tools can enhance their reliability in a lightweight manner, reducing reliance on complex static and dynamic analyses. Additionally, SLMs offer a more adaptable solution capable of generalizing across different refactoring types and programming languages, addressing key limitations of traditional approaches.

\end{abstract}

\section{Introduction}

Refactoring~\cite{Fowler-book-1999,Opdyke-PHD-1992} is the process of restructuring a program to improve its internal quality while preserving its observable behavior. The first refactoring tool was proposed by Roberts~\cite{Roberts-PHD-1999}. Integrated Development Environments (IDEs), such as \eclipse{} and \netbeans{}, provide automated support for various refactorings, making the process more accessible to developers. This automation is important for maintaining code quality and improving software maintainability over time.

However, ensuring that refactoring transformations preserve a program’s behavior remains a significant challenge~\cite{Schafer-PLPV-2009}. Despite extensive verification efforts, automated refactorings in IDEs frequently contain bugs that introduce compilation or runtime errors or alter program behavior. Several approaches have been proposed to mitigate these issues, often focusing on detecting errors introduced by refactoring implementations~\cite{Soares-TSE-2013,test-tools-fse07,steimann-ecoop}. For instance, \saferefactor{}~\cite{saferefactor-ieee} automatically generates and runs test cases to detect unintended behavioral changes. Other techniques address cases where a valid refactoring is unnecessarily blocked due to overly restrictive preconditions~\cite{Mongiovi-TSE-2018,Soares-ICSM-2011}. Implementing correct refactorings is inherently complex, as it requires defining and enforcing multiple preconditions while managing diverse transformations~\cite{Schafer-PLPV-2009,oliveira2019revisiting}. The difficulty of ensuring correctness in refactoring tools may explain why many developers still prefer manual refactorings~\cite{Tempero-ACM-2017}.

More recently, Large and Small Language Models (LLMs and SLMs) have emerged as a promising, cost-effective alternative for detecting software defects, offering efficient analysis on consumer-grade hardware~\cite{se-llms-2023,wang2024}. Compared to large-scale models such as \gpt{} and \claude{}, SLMs require fewer computational resources, making them easier to integrate into real-world development environments, including continuous integration pipelines and on-premise setups. Additionally, their reduced size allows for lower inference latency and more efficient fine-tuning, making them accessible to organizations with limited computational infrastructure. However, the effectiveness of SLMs in accurately identifying refactoring-related bugs remains an open question~\cite{DBLP:conf/fose-ws/FanGHLSYZ23}, to the best of our knowledge.

This study addresses this gap by evaluating the extent to which \slms{} can detect refactoring bugs in Java and Python. We assess the performance of widely used SLMs in identifying \bugs{} across 16 refactoring types reported in popular IDEs such as \eclipse{} and \netbeans{}. We are analyzing bugs related to code transformations rather than traditional bugs that occur within a single program. Our findings indicate that SLMs successfully detected 71.4\% of the reported bugs involving incorrect transformations (\textsc{Type I} bugs). Notably, the proprietary \ominihigh{} model achieved the highest detection rate, while the open-source \phimodel{} demonstrated comparable performance across both bug categories. However, identifying \textsc{Type II} bugs -- valid refactorings incorrectly blocked by IDEs -- remains a challenge, with models achieving a success rate of only 40\%{}. A mixture of experts approach could be a highly effective strategy for enhancing refactoring bug detection, leveraging the strengths of multiple models to improve accuracy and coverage.

This study enhances the understanding of how SLMs can be leveraged for software maintenance, identifying opportunities to improve automated refactoring validation tools. Unlike previous approaches that depend on predefined preconditions for specific refactoring types, our method has the potential to generalize across different refactorings and programming languages, increasing scalability and adaptability. Furthermore, integrating SLMs into IDEs using our setup, similar to Windsurf~\cite{windsurf} and Cursor~\cite{cursor} AI-powered IDEs, can enhance reliability and automation, potentially reducing dependence on manual refactorings, which remain widespread despite the availability of automated tools~\cite{Tempero-ACM-2017}. A cost-effective approach would be to use \phimodel{} for most cases and leverage \ominihigh{} only when human reviewers are uncertain, optimizing resource allocation while maintaining high detection accuracy. 

This article is structured as follows. Section~\ref{sec:example} introduces a motivating example. Sections~\ref{sec:methodology} and~\ref{sec:results} detail our methodology and present our findings, respectively. Section~\ref{sec:discussion} provides an in-depth discussion of the results. Section~\ref{sec:relwork} examines related work, and Section~\ref{sec:conclusion} summarizes our conclusions and outlines future research directions.

\section{Motivating Example}
\label{sec:example}

To ensure that a refactoring transformation preserves program correctness, it must satisfy a set of \textit{preconditions}. These preconditions guarantee that the resulting program compiles, maintains or improves its quality, and preserves the original program’s behavior. For example, in the \textit{Extract Class} refactoring~\cite{Fowler-book-1999}, the newly introduced class must not have a name that conflicts with an existing class to avoid compilation errors.
In practice, developers can refactor code manually, which is error-prone and time-consuming, or use automated tools that support various refactorings, such as \vscode{}~\cite{vscode}, \intellij{}~\cite{intellij}, \eclipse{}~\cite{eclipse}, and \netbeans{}~\cite{netbeans}. 

Currently, verifying behavior preservation is commonly done by compiling the refactored program and running its test suite. However, existing tools are not entirely reliable. They may introduce defects ranging from compilation errors, which are relatively straightforward to detect, to subtle behavioral changes, which are significantly harder to identify. This issue arises because many refactoring implementations do not formally define all necessary preconditions to ensure behavior preservation. In practice, most refactoring tools only check a subset of these preconditions.

\subsection{Refactoring Implementations with Incomplete Preconditions (\textsc{Type I} Bugs)}

To illustrate the issue of insufficient preconditions in refactoring implementations (\textsc{Type I} bugs), consider the program in Listing~\ref{exampleOWCs}, which declares three classes. Suppose the \textit{Push Down Method} refactoring is applied to move method \texttt{B.m} to \texttt{C} using \eclipse{} JDT. This transformation yields the refactored program shown in Listing~\ref{exampleOWCt}.

\begin{figure*}
\centering
\begin{minipage}{2.3in}
\begin{lstlisting}[basicstyle=\footnotesize,language=Java, label=exampleOWCs, caption={Original program.}]
public class A {
  public int k() {
    return 10;
  }
}
public class B extends A {
  public int k() {
    return 20;
  }
  public int m() {
    return super.k();
  }
}
public class C extends B {
  public static void main(String[] args) {
    C c = new C();
    System.out.println(c.m());
  }
}
\end{lstlisting}
\end{minipage}
\begin{minipage}{0.9in}
\end{minipage}
\begin{minipage}{2.3in}
 \begin{lstlisting}[basicstyle=\footnotesize,language=Java, label=exampleOWCt, caption={Refactored program.}]
public class A {
  public int k() {
    return 10;
  }
}
public class B extends A {
  public int k() {
    return 20;
  }
}
public class C extends B {
  public int m() {
    return super.k();
  }
  public static void main(String[] args) {
    C c = new C();
    System.out.println(c.m());
  }
}
\end{lstlisting}
\end{minipage}
\caption{Applying Push Down Method \texttt{B.m} to \texttt{C} using \eclipse{} JDT introduces a behavioral change.}
\label{fig:exampleOWC}
\end{figure*}

Executing \texttt{C.main} before refactoring produces \texttt{10}, whereas after refactoring, it produces \texttt{20}. This change occurs because \eclipse{} does not modify the call to \texttt{super.k()} when pushing down a method to a subclass, leading to an unintended behavior change.

Although this transformation is applied to a small program, evidence suggests that similar issues may occur in real-world software projects, underscoring the importance of studying these transformations~\cite{Gligoric-ecoop-13,soares-jss-2013}. For instance, Gligoric et al.~\cite{Gligoric-ecoop-13} applied refactorings to five large open-source Java projects—\textit{JPF}, \textit{JUnit}, \textit{Apache Log4j}, \textit{Apache Lucene}, and \textit{Apache Commons Math} -- and identified 77 refactoring -- related bugs in \eclipse{}, many of which resemble the issue illustrated in Figure~\ref{fig:exampleOWC}. 
Moreover, these issues are not isolated cases; refactoring tools across multiple IDEs, including \intellij{}, \vscode{}, \eclipse{}, and \netbeans{}, exhibit similar errors due to the inherent complexity of specifying and enforcing all necessary preconditions to ensure behavior preservation. Formalizing refactoring correctness using formal semantics is challenging~\cite{Schafer-PLPV-2009}, leading developers to implement only a subset of the required preconditions. This limitation may hinder the widespread adoption of automated refactoring tools, as developers may question their reliability and favor manual refactoring to mitigate potential risks~\cite{Tempero-ACM-2017}.

\subsection{Refactoring Implementations Restricting Valid Transformations (\textsc{Type II} Bugs)}

In contrast to \textsc{Type I} bugs, where refactorings apply but introduce defects, some tools impose overly strict preconditions, unnecessarily blocking valid transformations (\textsc{Type II} bugs). Consider Listing~\ref{exampleOSCs2}, which illustrates a portion of a program that manages database queries while supporting two database versions. Each version is implemented in a separate class: \texttt{QueryV1} (for database version 1) and \texttt{QueryV2} (for database version 2). These classes allow client code to switch between database versions seamlessly. Both \texttt{QueryV1} and \texttt{QueryV2} extend a common abstract class, \texttt{Query}, which defines an abstract method, \texttt{createQuery}. Each subclass provides its own implementation of this method. The queries generated by \texttt{createQuery} are executed by the \texttt{doQuery} method. However, this method is duplicated in both subclasses, \texttt{QueryV1} and \texttt{QueryV2}, introducing redundancy in the codebase.

\begin{figure*}
\centering
\begin{minipage}{2.5in}
\begin{lstlisting}[basicstyle=\footnotesize,language=Java, label=exampleOSCs2, caption={Original program.}]
public abstract class Query {
  protected abstract SDQuery createQuery();
}
public class QueryV1 extends Query {
  public void doQuery () {
    SDQuery sd = createQuery ();
    ...
  }
  protected SDQuery createQuery () {
    ...
  }
}
public class QueryV2 extends Query {
  public void doQuery () {
    SDQuery sd = createQuery ();
    ...
  }
  protected SDQuery createQuery () {
    ...
  }
}
\end{lstlisting}
\end{minipage}
\begin{minipage}{0.9in}
\end{minipage}
\begin{minipage}{2.7in}
 \begin{lstlisting}[basicstyle=\footnotesize,language=Java, label=exampleOSCt2, caption={A possible transformed program.}]



 
public abstract class Query {
  protected abstract SDQuery createQuery();
  public void doQuery () {
    SDQuery sd = createQuery ();
    ...
  }
}
public class QueryV1 extends Query {
  protected SDQuery createQuery () {
    ...
  }
}
public class QueryV2 extends Query {
  protected SDQuery createQuery () {
    ...
  }
}
\end{lstlisting}
\end{minipage}
\caption{The Pull Up Method \texttt{doQuery} refactoring is not allowed by \eclipse{} JDT due to overly strong preconditions in its implementation.}
\label{fig:exampleOSC2}
\end{figure*}

We can eliminate duplication by pulling up the \texttt{doQuery} method. However, when applying this refactoring using \eclipse{} JDT 2.1, the tool issues a warning stating that \texttt{doQuery} does not have access to \texttt{createQuery}. This warning results from a precondition check that ensures the pulled-up method retains access to all of its called methods after the transformation. However, \texttt{createQuery} is already defined as an abstract method in the \texttt{Query} class, meaning this precondition is unnecessarily strict.

This issue was reported in \eclipse{}'s bug tracker.\footnote{\url{https://bugs.eclipse.org/bugs/show_bug.cgi?id=39896}} Kerievsky encountered it while developing a refactoring mechanic to introduce the Factory Method pattern for his book~\cite{Kerievsky-book-2004}. He argued that ``there should be no warnings as the transformation is harmless and correct.'' The \eclipse{} developers later fixed the bug, and Listing~\ref{exampleOSCt2} illustrates the correct refactoring applied using \eclipse{} JDT 4.5.

This case highlights a broader concern among \eclipse{} developers about identifying overly strong preconditions in refactoring implementations. Mongiovi et al.~\cite{Mongiovi-TSE-2018} analyzed 2,559 assertions in \eclipse{}'s test suite, and found that 32\% were related to detecting \textsc{Type II} bugs, further underscoring the importance of refining refactoring constraints to avoid unnecessary restrictions.

\subsection{Improving the Reliability of Refactoring Tools}

To address these issues, previous research has proposed several techniques to test and improve the correctness of refactoring implementations~\cite{Soares-TSE-2013,test-tools-fse07,steimann-ecoop,mongiovi-scp-2014,dong-icse-2025}. For example, prior work~\cite{Soares-TSE-2013,saferefactor-ieee,Mongiovi-TSE-2018} successfully identified the bug illustrated in Figures~\ref{fig:exampleOWC} and~\ref{fig:exampleOSC2}. While these contributions are valuable, existing approaches typically rely on intricate static and dynamic analyses and demand significant manual effort to define and validate the preconditions for each refactoring type. As a result, these methods tend to be time-consuming and challenging to scale, particularly as new refactorings emerge. Moreover, adapting such techniques to different programming languages is non-trivial due to their dependence on language-specific syntax and semantics.

Recent work has explored the use of large language models (LLMs) to support various refactoring tasks. For instance, some approaches automate Move Method recommendations by combining deep learning and LLM-generated insights~\cite{ZHANG2025121753}, perform variable renaming using pre-trained models such as RefBERT~\cite{DBLP:conf/issta/LiuWWXWLJ23}, or simplify Python code through prompt-based rewriting~\cite{DBLP:conf/apsec/ShirafujiOSMW23}. Other studies propose prompt engineering strategies for applying refactorings with ChatGPT~\cite{white2023chatgptpromptpatternsimproving}, investigate how developers interact with LLMs during refactoring~\cite{DBLP:conf/msr/AlOmarVMNO24}, or evaluate LLM-based tools designed to recommend Extract Method refactoring~\cite{DBLP:conf/sigsoft/PomianBDKBSBD24}. More recently, Dong et al.~\cite{dong-icse-2025} introduced a ChatGPT-based approach that automatically generates test programs for refactoring engines by leveraging a feature library constructed from bug reports and test cases for seven types of Java refactorings. Their approach uncovered new \textsc{Type I} bugs across.

Despite these promising directions, existing work does not offer a general approach for verifying the correctness of arbitrary refactoring transformations performed by developers using Small Language Models (SLMs), applicable across different refactoring types and programming languages. In particular, no prior method systematically tackles the detection of \textsc{Type II} bugs using foundation models.

\section{Methodology}
\label{sec:methodology}

This section details the methodology employed in our study.

\subsection{Research Questions}

The objective of this study is to evaluate the effectiveness of Small Language Models (SLMs) in detecting two types of refactoring bugs reported in Integrated Development Environments (IDEs). The first type (\textsc{Type I}) occurs when an IDE incorrectly applies a refactoring, leading to compilation errors, runtime failures, or unintended behavioral changes in the program. The second type (\textsc{Type II}) consists of cases where an IDE unnecessarily prevents a valid refactoring from being applied.

\begin{description}
    \item[RQ$_{1}$] How effectively can SLMs detect incorrect refactoring transformations (\textsc{Type I})?  
    \item[RQ$_{2}$] Can SLMs successfully apply refactoring transformations that IDEs unnecessarily prevent (\textsc{Type II})?  
\end{description}

\subsection{Planning}

\slms{} were chosen because they represent a diverse set of state-of-the-art models with varying architectures, parameter sizes, and optimization techniques, allowing for a comprehensive evaluation of their capabilities in refactoring tasks. Additionally, they include models known for reasoning abilities (\phimodel{}, \deepseek{}), code-specific optimizations (\omini{}, \ominihigh{}), and general-purpose performance (\llama{}, \mistral{}, \gemma{}, \gemmaNew{}), ensuring a balanced comparison across different approaches to handling code transformations and bug detection.

We evaluate \bugs{} refactoring bugs reported in IDEs, such as \eclipse{} and \netbeans{}, with up to 61 lines of code (LOC). All programs in our dataset are small and contain only a single instance of a refactoring application. 
Some of these bugs may represent edge scenarios, but studying them is essential to assess the extent to which SLMs can handle complex and challenging transformations. Similar refactoring-related bugs have been identified in real-world software projects~\cite{Gligoric-ecoop-13}.

For \textsc{Type I} bugs, we analyze \bugsCE{} transformations that introduce compilation or runtime errors and \bugsBC{} transformations that cause unintended behavioral changes. We provide both the original code and the incorrectly generated code resulting from the refactoring implementation. These bugs span \refactoringTypesOWC{} refactoring types in Java and Python.
For \textsc{Type II} bugs, we evaluate 30 transformations across \refactoringTypesOSC{} refactoring types in Java. These transformations represent cases where IDEs such as \eclipse{} and \netbeans{} unnecessarily block valid refactorings. The transformations are applied to programs with up to 24 lines of code (LOC). In this scenario, we provide the original code along with the refactoring parameters. At least one refactoring implementation fails to correctly apply the transformation.

All \bugs{} bugs in our dataset can be identified using automated techniques and subsequently validated manually. For instance, all Java-related errors and behavioral changes can be detected using \saferefactor{}~\cite{saferefactor-ieee,mongiovi-scp-2014}. Similarly, all Python errors can be identified by running PyRe~\cite{pyre}, a static type checker for Python, on the transformed program. Additionally, all \textsc{Type II} bugs were originally detected using prior techniques~\cite{Soares-ICSM-2011,Mongiovi-TSE-2018}, which systematically identify unnecessary refactoring preconditions in IDEs.

\subsection{Model Evaluation and Execution}

To assess the selected SLMs, we employ the zero-shot prompting technique~\cite{prompts,prompt-techniques,zero-shot-prompt}, where the model is required to answer a question without prior explicit training on that specific task. The models evaluated in this study include \slms{}, developed by Meta~\cite{llama}, Google~\cite{gemma2}, DeepSeek~\cite{deepseek-r1}, Microsoft~\cite{phi-4}, Mistral~\cite{mistral}, and OpenAI~\cite{o-mini}.

Except for \omini{}, which was accessed via its API, all models were executed locally using the Ollama framework in Python on a MacBook Pro M3 with 18GB of RAM (February and March 2025). We specified the model name and base URL, using the default configurations provided by the LangChain Ollama API~\cite{ollama-setup}. For \omini{}, we used its Python API with the default settings, passing only the API key~\cite{openai-api}. Additionally, \ominihigh{} was evaluated via its web interface with its default configuration. 

The \textit{pass@\(k\)} metric is a widely adopted evaluation measure in the context of code generation and model correctness~\cite{pass-k}. It estimates the probability that a model produces at least one correct output within \(k\) generated attempts for a given task. This metric is particularly useful for evaluating the functional reliability of foundation models in scenarios where multiple attempts are allowed.
Formally, as shown in Equation~\ref{eq:passatk}, we compute the proportion of dataset elements for which the model produces at least one correct output within the \(k\) attempts. Let \(n\) denote the total number of elements in the dataset, and let \(c_{i,k}\) represent the number of correct outputs among the \(k\) attempts for the \(i\)-th element. The indicator function \(\delta_{i,k}\) returns 1 if at least one of the \(k\) outputs is correct (\(c_{i,k} \geq 1\)), and 0 otherwise. Averaging \(\delta_{i,k}\) across all \(n\) elements yields the final pass@\(k\) score, expressed as a percentage.

\begin{equation}
\text{pass@}k = 
\left( \frac{1}{n} \sum_{i=1}^{n} 
\delta_{i,k} 
\right) \times 100\%,
\quad
\delta_{i,k} =
\begin{cases}
1, & \text{if } c_{i,k} \geq 1 \\
0, & \text{otherwise}
\end{cases}
\label{eq:passatk}
\end{equation}

Complementing this, the \textit{Consistency} metric, defined in Equation~\ref{eq:semantic-consistency}, captures how stable or deterministic a model's behavior is when evaluated multiple times on the same input. Specifically, it measures the average proportion of correct responses over \(k\) attempts for each dataset element.
Given a dataset with \(n\) elements, and assuming \(k\) independent attempts per element, let \(c_{i,k}\) again denote the number of correct outputs for the \(i\)-th element. The metric calculates the ratio \(c_{i,k}/k\) for each element, then averages these values across all elements and expresses the result as a percentage.

\begin{equation}
\text{Consistency@}k = 
\left( \frac{1}{n} \sum_{i=1}^{n} 
\frac{c_{i,k}}{k} 
\right) \times 100\%
\label{eq:semantic-consistency}
\end{equation}

While pass@\(k\) reflects whether a model can eventually produce a correct output within a limited number of attempts, the Consistency metric provides a finer-grained view of how often the model succeeds across all attempts. Evaluating both metrics offers complementary insights into a model’s reliability: high pass@\(k\) indicates potential correctness, whereas high consistency reflects robustness and predictability in repeated executions. These metrics are particularly important when assessing foundation models for tasks where determinism and reproducibility are critical.

We manually analyzed the responses of all SLMs and compared their results against a baseline for \bugs{} bugs. In some cases, the SLM correctly identified that a transformation was incorrect but provided an inaccurate explanation. In such instances, we classified the response as incorrect, as a valid explanation is essential for effectively diagnosing refactoring bugs.
For \textsc{Type I} bugs, we manually reviewed cases where the model indicated that a refactoring was incorrect. Our baseline contains the primary reasons why a transformation is invalid. For Java transformations, these issues typically result in either compilation errors or behavioral changes. For Python, they may lead to compilation or runtime errors. If a transformation resulted in a compilation error, but the model incorrectly described it as solely affecting behavior without acknowledging the compilation failure, we classified the response as incorrect.
For \textsc{Type II} bugs, when a model produced an output, we first compiled the resulting program using JDK 23. If the program failed to compile, we classified the response as incorrect. If it compiled successfully, we manually verified whether the transformation adhered to the expected refactoring mechanics~\cite{Fowler-book-1999}. For example, in the Pull Up Method refactoring, we ensured that the method was correctly removed from the subclass and properly declared in the parent class.

\subsection{Prompts Used for Evaluation}

For \textsc{Type I} bugs, we prompt the model to determine whether two given Java programs exhibit identical observable behavior after refactoring. The following prompt was used:

\begin{mdframed}[backgroundcolor=cyan!5, linecolor=black, linewidth=0.5pt]
\footnotesize
\noindent You are a coding assistant specializing in refactoring, with 30 years of experience. \\
\noindent Consider the following initial program: \\
\texttt{code1} \\
After applying a refactoring, the resulting program is: \\
\texttt{code2} \\
\noindent Your task is to verify two conditions. First, does the resulting program compile successfully? Second, does the transformation preserve observable behavior? If the refactoring satisfies both conditions, respond with EXACTLY ``YES'' on the first line. If not, respond with EXACTLY ``NO'' on the first line, then explain why.
\end{mdframed}

In the prompt, \texttt{code1} and \texttt{code2} represent the code before and after the refactoring, respectively. For example, in Section~\ref{sec:example}, \texttt{code1} and \texttt{code2} correspond to the programs in Listings~\ref{exampleOWCs} and~\ref{exampleOWCt}, respectively.
This prompt employs role prompting, explicit instruction, binary constraints, format enforcement, and conditional explanations, ensuring a structured response for evaluating refactoring correctness~\cite{prompt-techniques,prompts}.

For \textsc{Type II} bugs, we prompt the model to determine whether a specific refactoring transformation should be applied. The following prompt was used:

\begin{mdframed}[backgroundcolor=cyan!5, linecolor=black, linewidth=0.5pt]
\footnotesize
\noindent You are a coding assistant specializing in refactoring, with 30 years of experience.  \\
Can I apply the refactoring \texttt{name} to \texttt{params}? \\
\texttt{code} \\
\noindent If the refactoring can be applied, respond with EXACTLY ``YES'' on the first line, followed by the refactored program. If not, respond with EXACTLY ``NO'' on the first line, then explain why.
\end{mdframed}

In this prompt, \texttt{name} represents the type of refactoring, while \texttt{params} specifies the transformation parameters. For instance, consider Listing~\ref{exampleOSCs} (\texttt{code}) in Section~\ref{sec:example}. Here, \texttt{name} corresponds to the \textit{Push Down Field} refactoring, while \texttt{params} is defined as ``push down \texttt{A.f} to class \texttt{C}.''

\section{Results}
\label{sec:results}

\begin{figure*}[htp]
\centering

\begin{minipage}{\textwidth}
\centering
\begin{subfigure}{0.48\linewidth}
\centering
\begin{tikzpicture}
\begin{axis}[
    ybar,
    bar width=0.3cm,
    width=\linewidth,
    height=4.5cm,
    ylabel={\%},
    ymin=0, ymax=110,
    symbolic x coords={llama3.2:3B,gemma2:9B,deepseek-r1:14B,phi4:14B,mistral,o1-mini},
    xtick=data,
    x tick label style={rotate=30,anchor=east},
    nodes near coords,
    nodes near coords align={vertical},
]
\addplot coordinates {(llama3.2:3B,41.4) (gemma2:9B,40) (deepseek-r1:14B,60) (phi4:14B,71.4) (mistral,2.9) (o1-mini,70)};
\end{axis}
\end{tikzpicture}
\caption{Type I: Java and Python}
\label{fig:subfig-a}
\end{subfigure}
\hfill
\begin{subfigure}{0.48\linewidth}
\centering
\begin{tikzpicture}
\begin{axis}[
    ybar,
    bar width=0.3cm,
    width=\linewidth,
    height=4.5cm,
    ylabel={\%},
    ymin=0, ymax=110,
    symbolic x coords={llama3.2:3B,gemma2:9B,deepseek-r1:14B,phi4:14B,mistral,o1-mini},
    xtick=data,
    x tick label style={rotate=30,anchor=east},
    nodes near coords,
    nodes near coords align={vertical},
]
\addplot coordinates {(llama3.2:3B,52.6) (gemma2:9B,57.9) (deepseek-r1:14B,68.4) (phi4:14B,78.9) (mistral,5.3) (o1-mini,84.2)};
\end{axis}
\end{tikzpicture}
\caption{Type I: Behavioral Changes in Java}
\label{fig:subfig-b}
\end{subfigure}
\end{minipage}

\vspace{1em} 

\begin{minipage}{\textwidth}
\centering
\begin{subfigure}{0.48\linewidth}
\centering
\begin{tikzpicture}
\begin{axis}[
    ybar,
    bar width=0.3cm,
    width=\linewidth,
    height=4.5cm,
    ylabel={\%},
    ymin=0, ymax=110,
    symbolic x coords={llama3.2:3B,gemma2:9B,deepseek-r1:14B,phi4:14B,mistral,o1-mini},
    xtick=data,
    x tick label style={rotate=30,anchor=east},
    nodes near coords,
    nodes near coords align={vertical},
]
\addplot coordinates {(llama3.2:3B,34.8) (gemma2:9B,30.4) (deepseek-r1:14B,73.9) (phi4:14B,69.6) (mistral,4.3) (o1-mini,69.6)};
\end{axis}
\end{tikzpicture}
\caption{Type I: Java Errors}
\label{fig:subfig-c}
\end{subfigure}
\hfill
\begin{subfigure}{0.48\linewidth}
\centering
\begin{tikzpicture}
\begin{axis}[
    ybar,
    bar width=0.3cm,
    width=\linewidth,
    height=4.5cm,
    ylabel={\%},
    ymin=0, ymax=110,
    symbolic x coords={llama3.2:3B,gemma2:9B,deepseek-r1:14B,phi4:14B,mistral,o1-mini},
    xtick=data,
    x tick label style={rotate=30,anchor=east},
    nodes near coords,
    nodes near coords align={vertical},
]
\addplot coordinates {(llama3.2:3B,39.3) (gemma2:9B,35.7) (deepseek-r1:14B,46.4) (phi4:14B,67.9) (mistral,0) (o1-mini,64.3)};
\end{axis}
\end{tikzpicture}
\caption{Type I: Python Errors}
\label{fig:subfig-d}
\end{subfigure}
\end{minipage}

\vspace{1em} 

\begin{minipage}{\textwidth}
\centering
\begin{subfigure}{0.48\linewidth}
\centering
\begin{tikzpicture}
\begin{axis}[
    ybar,
    bar width=0.3cm,
    width=\linewidth,
    height=4.5cm,
    ylabel={\%},
    ymin=0, ymax=110,
    symbolic x coords={llama3.2:3B,gemma2:9B,deepseek-r1:14B,phi4:14B,mistral,o1-mini},
    xtick=data,
    x tick label style={rotate=30,anchor=east},
    nodes near coords,
    nodes near coords align={vertical},
]
\addplot coordinates {(llama3.2:3B,32.9) (gemma2:9B,21.4) (deepseek-r1:14B,12.9) (phi4:14B,21.4) (mistral,0) (o1-mini,15.7)};
\end{axis}
\end{tikzpicture}
\caption{Type I: Incorrect Explanation}
\label{fig:subfig-e}
\end{subfigure}
\end{minipage}

\caption{Detection rates (\%) across multiple models for \textsc{Type I} bugs.}
\label{fig:resultsOWC}
\end{figure*}
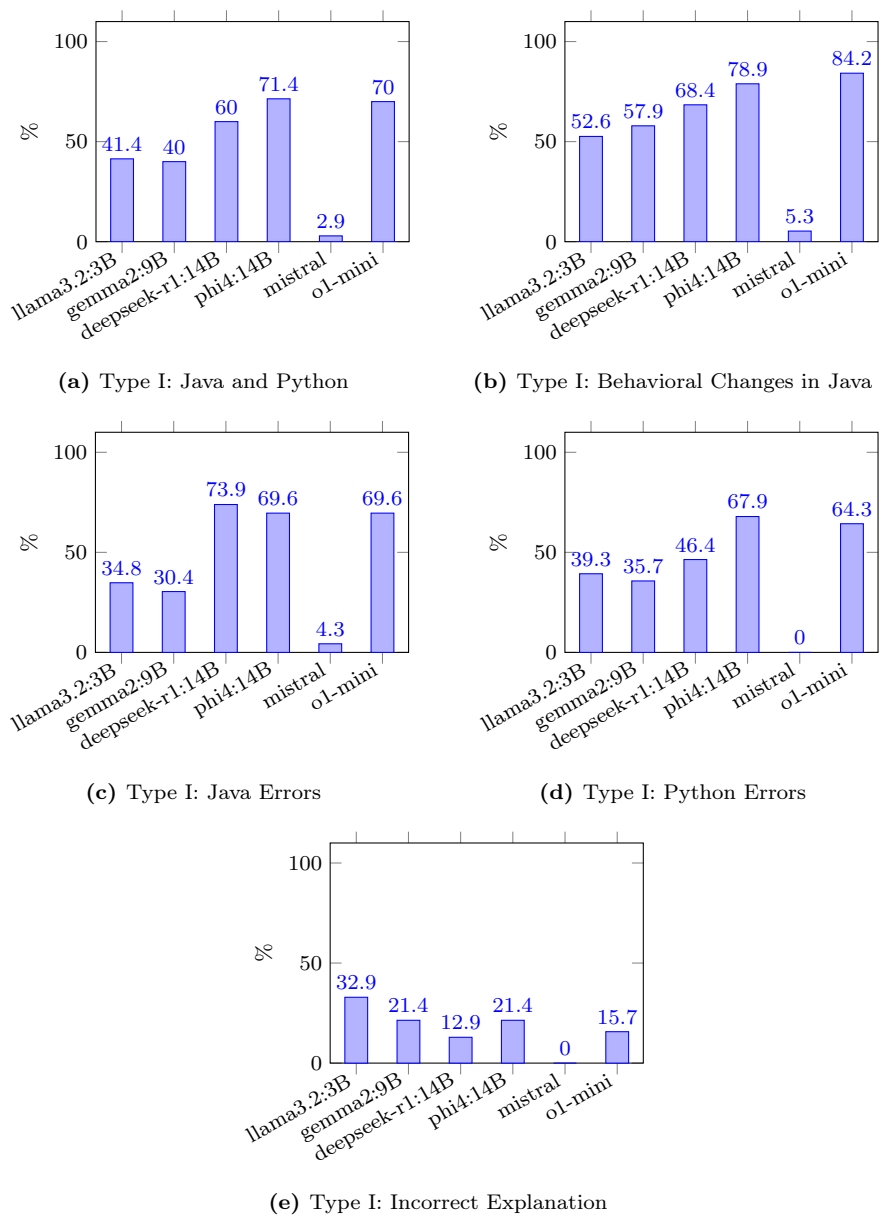

Overall, Figure~\ref{fig:subfig-a} shows that \phimodel{} achieves the highest detection rate (71.4\%) for \textsc{Type I} bugs across Java and Python, outperforming all other models, with \omini{} following closely at 70\%. In contrast, \mistral{} exhibits the weakest performance (2.9\%), detecting almost no refactoring bugs.
For \textsc{Type I} bugs, at least one model successfully detects all but one bug (1.4\%), which introduces an error in a Python program. Similarly, for \textsc{Type II} bugs, at least one model detects all but nine bugs (30\%). These results suggest that a mixture of experts approach -- leveraging multiple models -- may be an effective strategy for detecting the majority of refactoring bugs, as different models excel in different aspects of bug detection.

Figure~\ref{fig:subfig-b} breaks down \textsc{Type I} bugs in Java into Behavioral Changes and Compilation Errors. \omini{} achieves the highest performance, correctly detecting 84.2\% of behavioral changes, followed by \phimodel{} (78.9\%) and \deepseek{}. For compilation errors, \deepseek{} leads with 73.9\% detection, closely followed by \phimodel{} and \omini{} (see Figure~\ref{fig:subfig-c}). Meanwhile, \llama{} and \gemma{} show moderate performance, with detection rates ranging between 30.4\% and 57.9\%. In contrast, \mistral{} consistently underperforms, detecting only 5.3\% of behavioral changes and 4.3\% of Java compilation errors.

For \textsc{Type I} bugs in Python (Figure~\ref{fig:subfig-d}), the detection rates are lower compared to Java. \phimodel{} maintains the highest accuracy (67.9\%), followed by \omini{} (64.3\%). However, all models struggle more with Python errors, likely due to its dynamic typing system and the lack of static analysis checks. Notably, \mistral{} completely fails to detect any Python-related errors.

Figure~\ref{fig:subfig-e} presents an analysis of incorrect explanations given by the models for \textsc{Type I} bugs. Some models correctly identify whether a refactoring is incorrect but fail to provide an accurate justification. \llama{} has the highest rate of incorrect explanations (32.9\%), followed by \gemma{} (21.4\%) and \phimodel{} (21.4\%). This suggests that while models can identify refactoring issues, they often struggle to provide accurate explanations for the detected problems. As a result, developers must carefully review the model’s reasoning to verify its correctness and ensure informed decision-making.

\begin{figure*}[htp]
\centering

\begin{minipage}{\textwidth}
\centering
\begin{subfigure}{0.48\linewidth}
\centering
\begin{tikzpicture}
\begin{axis}[
    ybar,
    bar width=0.3cm,
    width=\linewidth,
    height=4.5cm,
    ylabel={\%},
    ymin=0, ymax=110,
    symbolic x coords={llama3.2:3B,gemma2:9B,deepseek-r1:14B,phi4:14B,mistral,o1-mini},
    xtick=data,
    x tick label style={rotate=30,anchor=east},
    nodes near coords,
    nodes near coords align={vertical},
]
\addplot coordinates {
    (llama3.2:3B,3.3)
    (gemma2:9B,36.7)
    (deepseek-r1:14B,36.7)
    (phi4:14B,30)  
    (mistral,16.7) 
    (o1-mini,36.7)
};
\end{axis}
\end{tikzpicture}
\caption{Type II: Java}
\label{fig:subfig-f}
\end{subfigure}
\hfill
\begin{subfigure}{0.48\linewidth}
\centering
\begin{tikzpicture}
\begin{axis}[
    ybar,
    bar width=0.3cm,
    width=\linewidth,
    height=4.5cm,
    ylabel={\%},
    ymin=0, ymax=110,
    symbolic x coords={llama3.2:3B,gemma2:9B,deepseek-r1:14B,phi4:14B,mistral,o1-mini},
    xtick=data,
    x tick label style={rotate=30,anchor=east},
    nodes near coords,
    nodes near coords align={vertical},
]
\addplot coordinates {
    (llama3.2:3B,96.7)
    (gemma2:9B,40)
    (deepseek-r1:14B,10)
    (phi4:14B,26.7)  
    (mistral,76.7) 
    (o1-mini,13.3)
};
\end{axis}
\end{tikzpicture}
\caption{Type II: Incorrect Explanation}
\label{fig:subfig-g}
\end{subfigure}
\end{minipage}

\caption{Detection rates (\%) across multiple models for \textsc{Type II} bugs.}
\label{fig:resultsOSC}
\end{figure*}
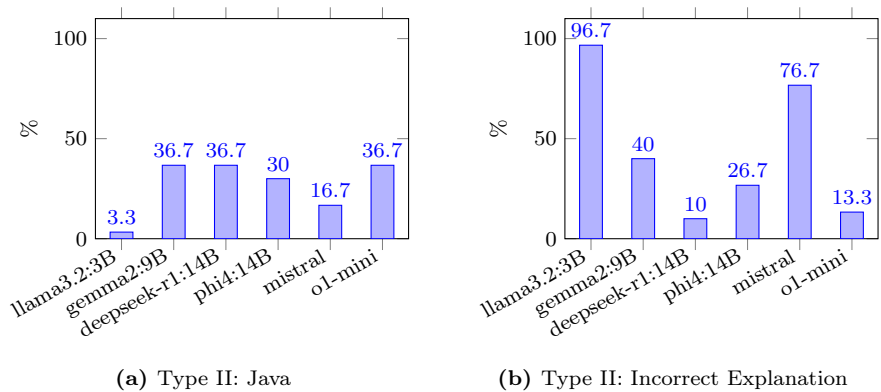

For \textsc{Type II} bugs—cases where IDEs unnecessarily block refactorings -- the detection rates drop considerably (Figure~\ref{fig:subfig-f}). \phimodel{}, \llama{}, \gemma{}, \deepseek{}, and \omini{} each reach a detection rate of 36.7\%, showing a more balanced performance across models. However, \mistral{} struggles significantly, detecting only 16.7\% of these cases. Figure~\ref{fig:subfig-g} further highlights incorrect explanations for \textsc{Type II} bugs. \llama{} has the highest rate of incorrect explanations (96.7\%), suggesting that while it attempts to justify refactoring decisions, it frequently provides incorrect reasoning.

In summary, \phimodel{} and \omini{} demonstrate the strongest performance in detecting both \textsc{Type I} and \textsc{Type II} refactoring bugs, consistently outperforming other models across various refactoring scenarios. While some models provide reasonable detection rates, their explanations often lack precision. Additionally, identifying unnecessarily blocked refactorings (\textsc{Type II}) remains a challenge for all models, suggesting the need for further refinements. 

\section{Discussion}
\label{sec:discussion}

Next we discuss the results of our work.

\subsection{\textsc{Type I} Bugs}

\textsc{Type I} bugs encompass multiple subtypes, each presenting distinct challenges. For instance, a transformation in Java may introduce a compilation error due to incorrect modifications of syntax or type constraints (see Figure~\ref{fig:subfig-c}). In one case, \phimodel{} failed to detect a compilation error introduced by the Rename Method refactoring in \netbeans{}. Specifically, in Java, \texttt{A.this} can only be used inside an inner (non-static) class within \texttt{A}, whereas \texttt{B} is a subclass, not an inner class. The correct fix would be to call \texttt{super.m("1")} instead. However, \phimodel{} did not identify this issue. Among all the models analyzed, only \ominihigh{} successfully detected the compilation error. Some SLMs, such as \llama{}, incorrectly stated that the resulting code compiles, even though the transformation introduces a behavioral change. SLMs do not fully grasp all the constraints required for a Java program to be correct. The accuracy in detecting this type of bug could be improved by integrating a compilation tool into SLMs, enabling them to validate code correctness more effectively.

In Python, certain errors only manifest at runtime, making static detection significantly more challenging (see Figure~\ref{fig:subfig-d}). Unlike Java, where type constraints are enforced at compile time, Python’s dynamic typing system allows type mismatches and incorrect method invocations to remain undetected until execution.
For instance, consider the Python code in Listing~\ref{exampleOWCs2}. Suppose we apply a Rename Variable refactoring to \texttt{text}, renaming it to the Python reserved keyword \texttt{continue} using Rope~\cite{rope}. This transformation introduces an error, yet \mistral{} and \phimodel{} failed to detect it, while the other models correctly identified that the refactored code in Listing~\ref{exampleOWCt2} is invalid.

\begin{figure*}
\centering
\begin{minipage}{2.3in}
\begin{lstlisting}[basicstyle=\footnotesize,language=Python, label=exampleOWCs2, caption={Original program.}]
  class B:
    def a(self) -> None:
      text = `hello'
\end{lstlisting}
\end{minipage}
\begin{minipage}{0.9in}
\end{minipage}
\begin{minipage}{2.3in}
 \begin{lstlisting}[basicstyle=\footnotesize,language=Python, label=exampleOWCt2, caption={Refactored program.}]
  class B:
    def a(self) -> None:
      continue = `hello'
\end{lstlisting}
\end{minipage}
\caption{Applying Rename Variable \texttt{text} to \texttt{continue} using \rope{} introduces an error.}
\label{fig:exampleOWC2}
\end{figure*}

Although this is a simple case, we also evaluated more complex refactoring bugs that introduce runtime errors in Python. Our dataset reveals a consistent performance drop across all models when identifying these runtime-specific errors. While tools such as PyRe~\cite{pyre} and Pytype~\cite{pytype} assist developers with static analysis, they remain limited in detecting errors that require deeper reasoning beyond type inference. Integrating these tools into SLMs could enhance their ability to detect a broader range of errors, improving the reliability of automated refactoring detection in dynamically typed languages like Python.

Another critical subtype involves transformations that produce a program that compiles but alters its behavior. These semantic inconsistencies are particularly difficult to identify~\cite{daniel-fse-07,Soares-TSE-2013}, as even comprehensive test suites may fail to detect subtle changes. Specialized tools mitigate this issue by automatically generating test cases for entities affected by the transformation, leveraging sophisticated static and dynamic analysis techniques~\cite{saferefactor-ieee,mongiovi-scp-2014}. Notably, \omini{} and \phimodel{} demonstrate strong performance in detecting behavioral inconsistencies, providing clear and accurate explanations of the observed behavioral changes.

For example, we evaluate the transformation presented in Figure~\ref{fig:exampleOWC}. \mistral{} incorrectly assumes that the \texttt{main} method will output \texttt{20} instead of \texttt{10} in Listing~\ref{exampleOWCs}. \llama{} provides a general but suboptimal explanation of the behavioral change. \gemma{} mistakenly states that the transformation is correct. In contrast, \omini{}, \phimodel{} and \deepseek{} correctly explain the behavioral change. For instance, the following is \deepseek{}'s explanation:
\begin{mdframed}[backgroundcolor=light-gray, linecolor=black, linewidth=0.5pt]
\textit{``... The refactoring changes the observable behavior of the program. In the initial code, when \texttt{C.c.m()} is called, it executes \texttt{B.m()}, which returns \texttt{A.k()} (\texttt{10}). After refactoring, \texttt{C.m()} overrides and returns \texttt{B.k()} (\texttt{20}), altering the output from \texttt{10} to \texttt{20}....''}
\end{mdframed}

In our study, SLMs determine whether a transformation is correct or not. Figure~\ref{fig:subfig-e} shows the percentage of cases where a model identifies a transformation as incorrect but fails to provide a supporting explanation. For instance, a model might claim that the transformed code introduces a compilation error when, in reality, the code compiles but exhibits different behavior. Through manual analysis, we found that \llama{} provided an inaccurate explanation in 32.9\% of the cases where it detected an incorrect transformation. In contrast, \mistral{} identified only 2 out of 70 bugs but correctly explained both cases. While this type of analysis is time-consuming, it is crucial for assessing the reliability of models in automated refactoring scenarios. Notably, all models provided incorrect explanations for some bugs, highlighting the continued necessity of human evaluation to ensure accurate assessments.

Across all subtypes, \omini{} and \phimodel{} demonstrated the highest accuracy in detecting compilation, runtime, and behavioral errors. The practical implications of these findings suggest that integrating high-performing models like \phimodel{} into IDEs or continuous integration pipelines could significantly enhance the reliability of automated refactoring tools. A cost-effective strategy would be to use \phimodel{} in most cases, relying on \ominihigh{} only when human reviewers are uncertain about the results, thereby optimizing resource usage. 

\subsection{\textsc{Type II} Bugs}

For this category of bugs, we focused on transformations applied to Java programs that are unnecessarily blocked by IDEs. Each refactoring operation can often be performed in multiple ways~\cite{oliveira2019revisiting}, with minor adjustments that may help satisfy certain refactoring preconditions~\cite{daniel-icse23}. However, accounting for all possible refactoring variations is a complex task for tool developers, leading to cases where valid transformations are incorrectly prevented~\cite{Mongiovi-TSE-2018,Soares-ICSM-2011}. Our dataset includes 30 instances of \textsc{Type II} bugs, each representing a scenario where a refactoring tool incorrectly blocked a valid transformation.

For example, we evaluate the code presented in Listing~\ref{exampleOSCs}, which includes the \texttt{A} class and its subclasses \texttt{B} and \texttt{C}. Both \texttt{A} and \texttt{B} contain the \texttt{f} field, while \texttt{B} also declares a test method that accesses \texttt{B.f}, producing the value \texttt{1}. When attempting to use \jrrt{}~\cite{Schafer-OOPSLA-2010} to apply the Push Down Field refactoring -- moving \texttt{A.f} to class \texttt{C} -- the tool rejects the transformation due to an overly strong precondition.

\omini{} was the only model that correctly applied the transformation and provided an accurate description. The resulting code is equivalent to the one in Listing~\ref{exampleOSCt}. \gemma{}, \deepseek{}, and \phimodel{} do not allow the transformation to be applied. Specifically, \phimodel{} struggles with visibility preconditions, while \gemma{} encounters challenges in resolving ambiguity. \llama{} and \mistral{} permit the transformation but generate incorrect code that fails to push down the field. \ominihigh{} explains that refactoring is not possible yet still produces a valid code.

\begin{figure*}
\centering
\begin{minipage}{1.8in}
\begin{lstlisting}[basicstyle=\footnotesize,language=Java, label=exampleOSCs, caption={Original program.}]
public class A {
  private int f = 0;
}
public class B extends A {
  protected int f = 1;
  public long t() {
    return f;
  }
}
public class C extends A {}
\end{lstlisting}
\end{minipage}
\begin{minipage}{0.9in}
\end{minipage}
\begin{minipage}{2.6in}
 \begin{lstlisting}[basicstyle=\footnotesize,language=Java, label=exampleOSCt, caption={A possible transformed program.}]
public class A {}
public class B extends A {
  protected int f = q;
  public long t() {
    return f;
  }
}
public class C extends A {
  private int f = 0;
}
\end{lstlisting}
\end{minipage}
\caption{The \textit{Push Down Field A.f} refactoring to class C is not allowed by JRRT~\cite{Schafer-OOPSLA-2010} due to overly strict preconditions in its implementation.}
\label{fig:exampleOSC}
\end{figure*}

Our results indicate that detecting \textsc{Type II} bugs remains a significant challenge for both open and proprietary SLMs (see Figure~\ref{fig:subfig-f}). The best-performing model correctly applied only 36.7\% of the transformations, while in several cases, the models suggested refactorings that introduced compilation errors or behavioral changes instead of correctly applying the intended transformation. This suggests that current SLMs struggle with understanding and reasoning about the underlying preconditions imposed by refactoring tools.

In our study, SLMs determine whether a transformation can be applied or not. Figure~\ref{fig:subfig-g} presents the percentage of cases where a model applies a refactoring but produces incorrect code. For instance, a model might generate uncompilable code, alter the program's behavior, or fail to correctly implement the refactoring mechanics~\cite{Fowler-book-1999}. \gemma{} incorrectly deleted a method in one case and, in another, failed to push down the expected method. Through manual analysis, we found that \llama{} produced incorrect code in 96.7\% of the cases, indicating a high failure rate. In contrast, \deepseek{} was unreliable in only 10\% of the cases, demonstrating significantly better accuracy. While this type of analysis is time-consuming, it is essential for evaluating the reliability of models in automated refactoring scenarios. Notably, all models introduced errors in some cases, reinforcing the importance of human oversight to ensure correctness and maintainability.

Addressing \textsc{Type II} bugs likely requires additional research and specialized techniques, building upon prior work on identifying overly restrictive refactoring constraints~\cite{Mongiovi-TSE-2018,Soares-ICSM-2011}. These findings emphasize that no single model or strategy is currently sufficient to handle all types of refactoring bugs effectively. Future work should explore hybrid approaches that combine SLMs with static analysis tools or customized fine-tuning to improve performance in these challenging cases.

\subsection{\ominihigh{}}

\phimodel{} performed on par with or outperformed the proprietary model \omini{} across our dataset for both \textsc{Type I} and \textsc{Type II} bugs (see Figures~\ref{fig:resultsOWC}~and~\ref{fig:resultsOSC}). The best-performing proprietary small model currently available is \ominihigh{}. To further evaluate its capabilities, we re-examined all \textsc{Type I} and \textsc{Type II} bugs that \omini{} failed to detect, re-running these undetected cases through \ominihigh{} using its web interface in February 2025. \ominihigh{} successfully identified 50\% of these previously undetected \textsc{Type I} bugs, achieving an overall detection rate of 84.3\% for \textsc{Type I} bugs (Figure~\ref{fig:subfig-a}), slightly surpassing \phimodel{}. For \textsc{Type II} bugs, it identified only 5.3\% of the previously undetected cases, resulting in an overall detection rate of 40\%. In some cases, \ominihigh{} states that it cannot apply a refactoring but still generates a transformed program.

All models, except for \omini{}, were executed locally on a MacBook Pro M3 with 18GB of RAM using Ollama, incurring no additional costs. Running \omini{} across all \bugs{} resulted in a total cost of USD 0.34, with USD 0.24 specifically allocated to \textsc{Type~I} bugs. This demonstrates that local execution of open-source models offers a cost-effective alternative for refactoring validation, while proprietary models requiring API access introduce additional expenses.

These results have important practical implications for integrating small language models into automated refactoring workflows. The strong performance of \phimodel{}, an open-source alternative, suggests that effective refactoring validation can be achieved without reliance on proprietary models, making the process more cost-effective and independent of external services. However, the superior performance of \ominihigh{} indicates that proprietary models still offer an edge in detection accuracy, making them valuable in high-assurance development environments where maximizing correctness is critical. These findings highlight a promising hybrid approach: open-source models can serve as cost-efficient first-pass validators, while proprietary models can be selectively used for high-confidence verification, optimizing both accuracy and operational costs in automated refactoring tools.

\subsection{Pass$@$\(k\) and Consistency$@$\(k\)}

Analyzing pass$@$\(k\) metrics and consistency$@$\(k\) is crucial for assessing both performance and reliability in classification or decision-making methods~\cite{pass-k}. In this framework, pass$@$\(k\) denotes the success rate when a system or algorithm is allowed up to \(k\) attempts. We specifically examine two types of refactoring bugs: \textsc{Type I} and \textsc{Type II}. Meanwhile, consistency gauges how stable the outcomes remain across multiple attempts.

\begin{figure}[htbp]
\centering

\begin{subfigure}[b]{0.45\textwidth}
\centering
\begin{tikzpicture}
\begin{axis}[
    width=\textwidth,
    height=0.6\textwidth,
    xlabel={\(k\)},
    ylabel={\%},
    xmin=1, xmax=10,
    xtick={1,2,3,4,5,6,7,8,9,10},
    grid=major,
    legend style={
        font=\scriptsize,        
        at={(0.5,1.02)},         
        anchor=south,
        legend columns=2         
    }
]

\addplot+[mark=*] coordinates {
    (1,87.1)  (2,91.4)  (3,94.3)  (4,97.1)  (5,98.6)
    (6,98.6)  (7,98.6)  (8,98.6)  (9,98.6)  (10,98.6)
};
\addlegendentry{TYPE I}

\addplot+[mark=*] coordinates {
    (1,53.3)  (2,63.3)  (3,76.7)  (4,86.7)  (5,86.7)
    (6,90.0)  (7,96.7)  (8,96.7)  (9,96.7)  (10,96.7)
};
\addlegendentry{TYPE II}

\addplot+[mark=*] coordinates {
    (1,76.96) (2,82.97) (3,89.02) (4,93.98) (5,95.03)
    (6,96.02) (7,98.03) (8,98.03) (9,98.03) (10,98.03)
};
\addlegendentry{TYPE I and II}

\end{axis}
\end{tikzpicture}
\caption{pass$@$\(k\) metrics}
\label{fig:k-pass}
\end{subfigure}
\hfill
\begin{subfigure}[b]{0.45\textwidth}
\centering
\begin{tikzpicture}
\begin{axis}[
    width=\textwidth,
    height=0.6\textwidth,
    xlabel={\(k\)},
    ylabel={\%},
    xmin=1, xmax=10,
    xtick={1,2,3,4,5,6,7,8,9,10},
    grid=major,
    legend style={
        font=\scriptsize, 
        at={(0.5,1.02)},
        anchor=south,
        legend columns=2
    }
]

\addplot+[mark=*] coordinates {
    (1,87.1)  (2,88.6)  (3,90.0)  (4,90.0)  (5,90.6)
    (6,91.0)  (7,91.0)  (8,90.5)  (9,90.2)  (10,90.0)
};
\addlegendentry{TYPE I}

\addplot+[mark=*] coordinates {
    (1,53.3)  (2,45.0)  (3,48.9)  (4,48.3)  (5,48.7)
    (6,49.4)  (7,51.4)  (8,52.9)  (9,53.7)  (10,53.0)
};
\addlegendentry{TYPE II}

\addplot+[mark=*] coordinates {
    (1,77.0)  (2,75.5)  (3,77.7)  (4,77.5)  (5,78.0)
    (6,78.5)  (7,79.1)  (8,79.3)  (9,79.2)  (10,78.9)
};
\addlegendentry{TYPE I and II}

\end{axis}
\end{tikzpicture}
\caption{Consistency metrics}
\label{fig:consistency}
\end{subfigure}

\caption{Evaluation of \phimodel{} performance using pass$@$\(k\) and consistency$@$\(k\) metrics for \(k\) values from 1 to 10 using default parameters.}
\label{fig:k-pass-consistency}
\end{figure}
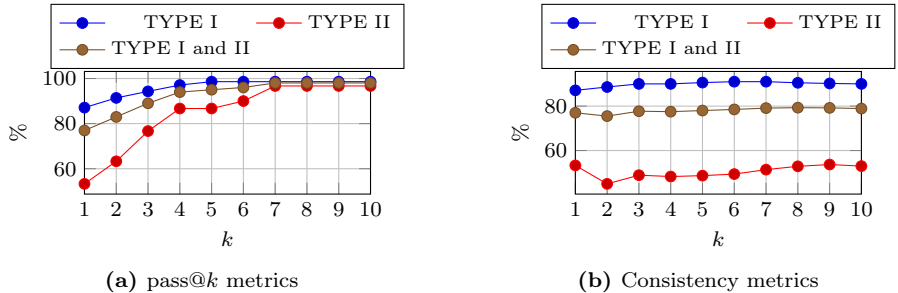

Using the same configuration presented before (Section~\ref{sec:methodology}), we evaluate the best open model (Section~\ref{sec:results}), \phimodel{}, focusing solely on its output. We do not manually evaluate its explanation. Our experiments consider both refactoring bug types over a range of \(k\) values from 1 to 10. 
In Figure~\ref{fig:k-pass}, the \textsc{TYPE I} line demonstrates good performance, quickly exceeding 90\% as \(k\) increases. By contrast, the \textsc{TYPE II} curve shows a more gradual progression, converging toward similarly high values by \(k=7\). The overall pass$@$\(k\) curve—an aggregate success rate—approaches near-perfect accuracy once \(k\) surpasses 5.

Figure~\ref{fig:consistency} presents the results for the consistency metrics. For both \textsc{Type I} and \textsc{Type II} bugs, consistency remains relatively stable, exhibiting only modest improvement at higher \(k\) values. On average, \phimodel{} achieves 78.07\% consistency across all evaluated \(k\). As with the pass$@$\(k\) results, \textsc{TYPE II} bugs pose a greater challenge, whereas \textsc{TYPE I} bugs are easier to detect and show more consistent outcomes in \phimodel{}’s predictions.

\subsection{Temperature}

Temperature is a key hyperparameter in foundation models that controls the randomness of generated outputs~\cite{temperature}. A lower temperature (closer to 0) makes the model more deterministic, favoring high-probability outputs and reducing variation. Conversely, a higher temperature increases randomness, allowing for more diverse responses at the cost of potential inconsistency. Evaluating different temperature values is crucial to understanding how a model balances accuracy and variability across different conditions. This analysis provides insights into whether a model maintains reliable performance under controlled randomness, which is particularly important for applications requiring stability in predictions, such as software engineering, legal reasoning, and medical diagnostics.

\begin{figure}[htbp]
    \centering
    \begin{tikzpicture}
        \begin{axis}[
            width=0.8\textwidth,
            height=0.5\textwidth,
            xlabel={Temperature},
            ylabel={\%},
            xmin=0, xmax=1,
            ymin=40, ymax=95,
            xtick={0,0.1,0.2,0.3,0.4,0.5,0.6,0.7,0.8,0.9,1},
            legend style={at={(1.02,0.5)}, anchor=west},
            grid=major
        ]
        
        \addplot+[mark=*] coordinates {
            (0,87.1) (0.1,90) (0.2,92.9) (0.3,90) (0.4,92.9)
            (0.5,92.9) (0.6,91.4) (0.7,87.1) (0.8,87.1) (0.9,88.6) (1,90)
        };
        \addlegendentry{TYPE I}
        
        \addplot+[mark=square*] coordinates {
            (0,56.7) (0.1,53.3) (0.2,53.3) (0.3,56.7) (0.4,50)
            (0.5,53.3) (0.6,63.3) (0.7,53.3) (0.8,46.7) (0.9,56.7) (1,43.3)
        };
        \addlegendentry{TYPE II}
        
        \addplot+[mark=triangle*] coordinates {
            (0,77.98) (0.1,78.99) (0.2,81.02) (0.3,80.01) (0.4,80.03)
            (0.5,81.02) (0.6,82.97) (0.7,76.96) (0.8,74.98) (0.9,79.03) (1,75.99)
        };
        \addlegendentry{TYPE I and II}
        
        \end{axis}
    \end{tikzpicture}
    \caption{Impact of temperature variation on the performance of \phimodel{}.}
    \label{fig:temperature}
\end{figure}
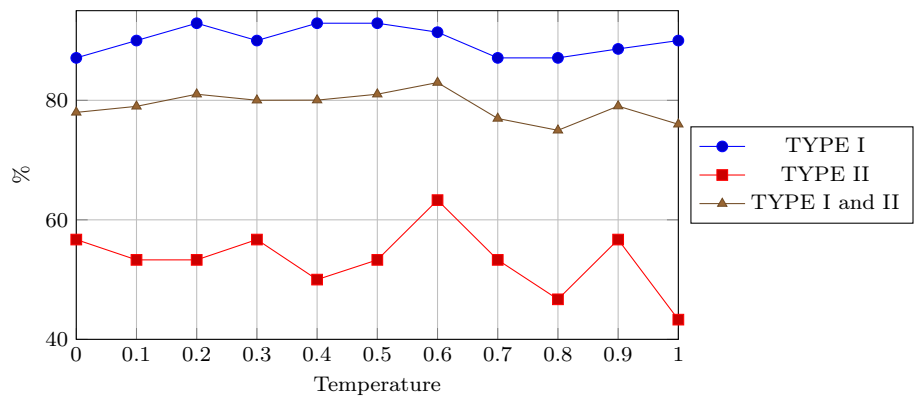

Figure~\ref{fig:temperature} presents the impact of temperature on the detection of \textsc{Type I} and \textsc{Type II} refactoring bugs, as well as the combined metric (\textsc{Type I and II}). The results show that \textsc{Type I} bugs are detected with consistently high accuracy, remaining above 87\%{} across all temperatures. This indicates that the model has a strong ability to recognize these cases, even when randomness increases. In contrast, \textsc{Type II} bug detection exhibits greater variability, with performance dropping to as low as 43.3\%{} at temperature 1.0. This suggests that the model struggles to maintain robust detection for \textsc{Type II} bugs as temperature increases, likely due to their more complex or ambiguous nature.

The combined metric (\textsc{Type I and II}) follows a pattern similar to \textsc{Type I}, but with slightly lower performance across all temperature values. The results indicate that a moderate temperature setting (around 0.5–0.6) achieves the best balance between accuracy and diversity, as this range maximizes the detection of both bug types while minimizing the drop in consistency. In Section~\ref{sec:methodology}, we set the temperature to 0.8. However, the results presented in Section~\ref{sec:results} could be enhanced by adjusting the temperature to a range of approximately 0.5–0.6, which balances accuracy and variability more effectively. For lower temperatures (0.0–0.3), the model maintains high accuracy but may be overly conservative, limiting its ability to generalize. For higher temperatures (0.7–1.0), detection performance becomes increasingly unstable, particularly for \textsc{Type II} bugs.

These findings highlight the trade-off between determinism and exploration in language models when applied to refactoring bug detection. Optimizing temperature selection is essential to ensuring that the model maintains high performance without sacrificing the ability to detect less common or more ambiguous cases. Future work may explore adaptive temperature mechanisms that adjust dynamically based on input complexity to further improve detection rates.

\subsection{Other Languages and Refactorings}

A key advantage of our approach is its flexibility, enabling its application across different programming languages and refactoring types. Unlike previous methods that are tailored to specific refactorings and rely on predefined preconditions~\cite{daniel-fse-07,Soares-TSE-2013,Mongiovi-TSE-2018,steimann-ecoop}, our SLM-based approach generalizes across various transformations without requiring manual adaptations for each case. The same prompt structure can be applied to evaluate different refactorings beyond those tested in this study, making it adaptable to evolving development environments and new refactoring techniques. This flexibility allows for seamless integration into refactoring tools for multiple languages, extending its benefits beyond Java and Python while reducing the need for language-specific validation mechanisms.

As a feasibility study, we evaluate our approach to detecting \textsc{Type I} bugs in refactoring implementations within \eclipse{} CDT~\cite{Gligoric-ecoop-13}. Gligoric et al. identified 43 bugs when applying refactorings to real-world C programs, including \textit{GMP}, \textit{libpng}, and \textit{zlib}.
To illustrate one such issue, consider a C program where the \texttt{main} function prints a number, as shown in Listing~\ref{exampleOWCsC}. After applying the Extract Function refactoring, the resulting program is presented in Listing~\ref{exampleOWCtC}. This bug was reported to developers.\footnote{\url{https://bugs.eclipse.org/bugs/show_bug.cgi?id=396344}} However, the refactored \texttt{newFunction} method has an incorrect return type -- it should return \texttt{const char*} instead of \texttt{const char}, leading to a compilation error.

\begin{figure*}
\centering
\begin{minipage}{2.3in}
\begin{lstlisting}[basicstyle=\footnotesize,language=C, label=exampleOWCsC, caption={Original program.}]



int main() {
  int a=0;
  printf("%d",a);
  return 0;
}
\end{lstlisting}
\end{minipage}
\begin{minipage}{0.9in}
\end{minipage}
\begin{minipage}{2.3in}
 \begin{lstlisting}[basicstyle=\footnotesize,language=C, label=exampleOWCtC, caption={Refactored program.}]
const char newFunction() {
  return "%d";
}
int main() {
  int a=0;
  printf(newFunction(), a);
  return 0;
}
\end{lstlisting}
\end{minipage}
\caption{Applying Extract Function using \eclipse{} CDT introduces a compilation error.}
\label{fig:exampleOWCC}
\end{figure*}

To assess this issue, we reused the same \textsc{Type I} bug detection prompt with \ominihigh{}, which correctly identified the error and produced the following message:
\begin{mdframed}[backgroundcolor=light-gray, linecolor=black, linewidth=0.5pt]
\textit{``... The refactored program does not compile because the function \texttt{newFunction} is declared to return a \texttt{const char} rather than a \texttt{const char*}. The return type mismatch causes a compilation error, so the transformation fails the first condition.}
\end{mdframed}
\phimodel{} also correctly identifies this transformation as incorrect, producing the following message:
\begin{mdframed}[backgroundcolor=light-gray, linecolor=black, linewidth=0.5pt]
\textit{``... The refactored program introduces a type mismatch. ... To fix this while maintaining correctness and behavior, the function should return "const char*" ...}
\end{mdframed}
As future work, we plan to extend our evaluation by analyzing a broader set of bugs across different programming languages.

\subsection{\gemmaNew{}}

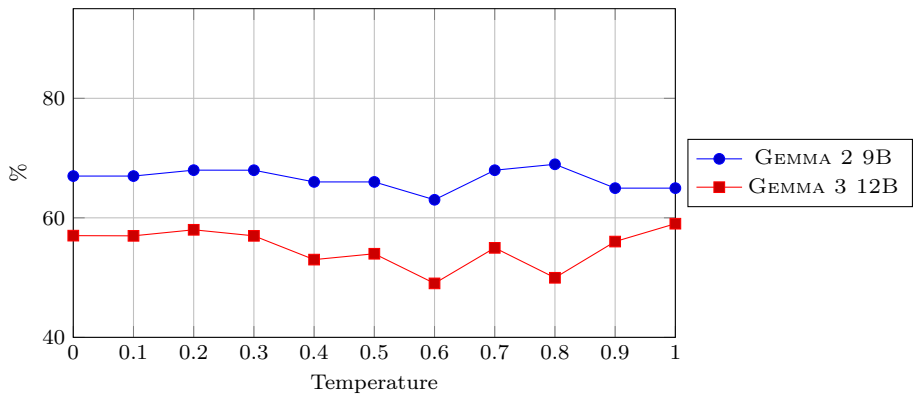
\begin{figure}[htbp]
    \centering
    \begin{tikzpicture}
        \begin{axis}[
            width=0.8\textwidth,
            height=0.5\textwidth,
            xlabel={Temperature},
            ylabel={\%},
            xmin=0, xmax=1,
            ymin=40, ymax=95,
            xtick={0,0.1,0.2,0.3,0.4,0.5,0.6,0.7,0.8,0.9,1},
            legend style={at={(1.02,0.5)}, anchor=west},
            grid=major
        ]
        
        \addplot+[mark=*] coordinates {
            (0, 66.99)  (0.1, 66.99)  (0.2, 67.98)  (0.3, 67.97)  (0.4, 66.01)  
            (0.5, 66.01)  (0.6, 63.02)  (0.7, 67.97)  (0.8, 68.96)  (0.9, 64.97)  (1, 64.97)
        };
        \addlegendentry{\gemma{}}
        
        \addplot+[mark=square*] coordinates {
            (0, 57.02)  (0.1, 56.99)  (0.2, 58)  (0.3, 56.99)  (0.4, 53.01)  
            (0.5, 53.99)  (0.6, 49.03)  (0.7, 54.98)  (0.8, 49.97)  (0.9, 56.03)  (1, 59.02)
        };
        \addlegendentry{\gemmaNew{}}
        
        \end{axis}
    \end{tikzpicture}
    \caption{Impact of temperature variation on the performance of \gemma{} and \gemmaNew{}.}
    \label{fig:temperature-gemma}
\end{figure}

In March 2025, \gemmaNew{} was released. In our study, we compared the performance of \gemma{} and \gemmaNew{}. Figure~\ref{fig:temperature-gemma} shows the impact of temperature variation on both models across a range from 0.0 to 1.0. We do not manually evaluate its explanation. \gemma{} demonstrates stable performance throughout, with scores consistently ranging from 66\% to 69\% and only minor fluctuations. In contrast, \gemmaNew{} exhibits greater variability, with a noticeable decline in performance between temperatures 0.3 and 0.7, reaching its lowest point around 0.6. Performance then improves as the temperature approaches 1.0. Overall, \gemma{} consistently outperforms \gemmaNew{}, and its stability across different temperature settings underscores its robustness under diverse sampling configurations.

\begin{figure}[htbp]
\centering

\begin{subfigure}[b]{0.45\textwidth}
\centering
\begin{tikzpicture}
\begin{axis}[
    width=\textwidth,
    height=0.6\textwidth,
    xlabel={\(k\)},
    ylabel={\%},
    xmin=1, xmax=10,
    xtick={1,2,3,4,5,6,7,8,9,10},
    grid=major,
    legend style={
        font=\scriptsize,        
        at={(0.5,1.02)},         
        anchor=south,
        legend columns=2         
    }
]

\addplot+[mark=*] coordinates {
(1, 66)  (2, 71.03)  (3, 73.01)  (4, 76.02)  (5, 76.02)  
(6, 76.02)  (7, 76.02)  (8, 77.04)  (9, 78.03)  (10, 78.03)
};
\addlegendentry{\gemma{}}

\addplot+[mark=*] coordinates {
(1, 57.99)  (2, 63.02)  (3, 70.01)  (4, 73.98)  (5, 76.01)  
(6, 77)  (7, 77.98)  (8, 79.03)  (9, 80.01)  (10, 80.99)
};
\addlegendentry{\gemmaNew{}}

\end{axis}
\end{tikzpicture}
\caption{pass$@$\(k\) metrics}
\label{fig:k-pass-gemma}
\end{subfigure}
\hfill
\begin{subfigure}[b]{0.45\textwidth}
\centering
\begin{tikzpicture}
\begin{axis}[
    width=\textwidth,
    height=0.6\textwidth,
    xlabel={\(k\)},
    ylabel={\%},
    xmin=1, xmax=10,
    xtick={1,2,3,4,5,6,7,8,9,10},
    grid=major,
    legend style={
        font=\scriptsize, 
        at={(0.5,1.02)},
        anchor=south,
        legend columns=2
    }
]

\addplot+[mark=*] coordinates {
(1, 66)  (2, 66.5)  (3, 67)  (4, 67.3)  (5, 67.2)  
(6, 66.8)  (7, 67.1)  (8, 67.4)  (9, 67.8)  (10, 67.8)
};
\addlegendentry{\gemma{}}

\addplot+[mark=*] coordinates {
(1, 58)  (2, 54.5)  (3, 53.3)  (4, 53.5)  (5, 53.2)  
(6, 53.8)  (7, 53.7)  (8, 54.1)  (9, 54.1)  (10, 53.8)
};
\addlegendentry{\gemmaNew{}}

\end{axis}
\end{tikzpicture}
\caption{Consistency metrics}
\label{fig:consistency-gemma}
\end{subfigure}

\caption{Evaluation of \gemma{} and \gemmaNew{} performance using pass$@$\(k\) and consistency metrics for \(k\) values from 1 to 10 using default parameters.}
\label{fig:k-pass-consistency-gemma}
\end{figure}
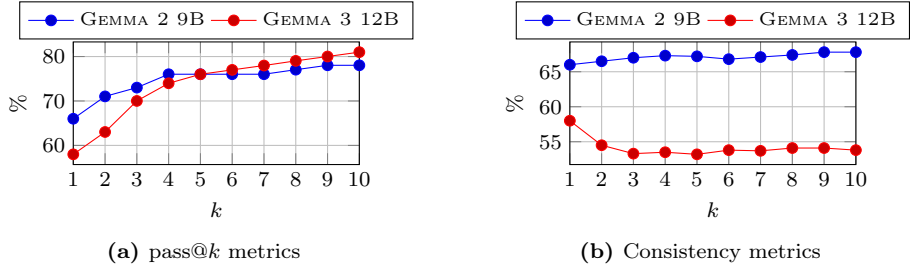

Figure~\ref{fig:k-pass-consistency-gemma} presents a comparative evaluation of \gemma{} and \gemmaNew{} across two different metrics: pass$@$\(k\) and consistency$@$\(k\), for values of \textit{k} ranging from 1 to 10. We evaluate both models using their optimal temperature settings: \gemma{} achieves its best performance at a temperature of 0.8, while \gemmaNew{} performs best at 1.0. In the pass$@$\(k\) plot (Figure~\ref{fig:k-pass-gemma}), both models show improved performance as \textit{k} increases, with \gemmaNew{} gradually catching up to \gemma{}. However, \gemma{} consistently outperforms \gemmaNew{}, especially for lower values of \textit{k}. In the consistency plot (Figure~\ref{fig:consistency-gemma}), \gemma{} maintains a high and stable performance above 65\%{} across all \textit{k} values, whereas \gemmaNew{} remains significantly lower, stabilizing around 55\%{}. These results suggest that \gemma{} is not only more consistent but also more effective at generating correct outputs across multiple attempts when compared to the newer \gemmaNew{} model under default settings.

\subsection{Metamorphic Testing}

A key threat to validity when evaluating foundation models is data contamination~\cite{sallou2024breaking}, where a model may have been exposed to similar or identical examples during training, leading to inflated performance results. To mitigate this risk, we employ metamorphic testing~\cite{metamorphic-testing,metamorphic-testing-2} to assess the robustness and reliability of the models. This technique generates new data samples by applying controlled metamorphic transformations to the original validation or testing data, ensuring that the semantic and behavioral equivalence of the modified code is preserved while altering its structural representation, such as the Abstract Syntax Tree (AST). The applied transformations include restructuring variables, parameters, classes and methods, as well as changing numbers, ensuring that the fundamental program logic remains intact. The primary objective is to evaluate the model’s resilience and its ability to detect refactoring issues consistently, even when minor syntactic variations are introduced.

To conduct this evaluation, we selected 10 instances of \textsc{Type I} bugs (compilation errors and behavioral changes) in Java that were successfully detected by \phimodel{}. Additionally, we evaluated 5 correct transformations applied by \phimodel{} on \textsc{Type II} bugs. using the same setup presented in Section~\ref{sec:methodology}. We then introduced controlled syntactic modifications to these examples before resubmitting them to the models. Specifically, we altered variable names, method names, class names, package names, and numeric parameters within the test code, ensuring variability while preserving the original program semantics.

\phimodel{} consistently identified all \textsc{Type I} transformations as incorrect. In 7 out of 10 cases, it correctly explained the reasoning behind the detection, missing two compilation errors and one behavioral change. For \textsc{Type II} bugs, it successfully applied the transformations and generated syntactically and semantically correct code in all cases.

These results reinforce confidence in \phimodel{}'s reliability as a tool for identifying and refactoring code issues. They suggest that its performance is based on an understanding of program behavior rather than mere reliance on exact textual patterns, making it more robust to variations in code structure.

\subsection{Outputs}

In our prompts, we instructed the models to follow a structured output format. For instance, for \textsc{Type II} bugs, we required models to respond with either ``YES'' on the first line, followed by the transformed Java program, or ``NO'' on the first line, followed by an explanation. \omini{}, \phimodel{}, and \gemma{} adhered to this structure most of the time, which facilitated automation in compiling Java programs when the response was ``YES.'' However, \deepseek{}, \mistral{} and \llama{} often deviated from this format, requiring additional manual processing.

For example, \deepseek{} provided the refactored code in fragmented sections interleaved with textual explanations, making it difficult to extract and compile automatically. Additionally, its responses tended to be excessively verbose, including unnecessary reasoning steps and occasional contradictions. In contrast, \omini{} models were notably more concise, providing clear and structured explanations for why a transformation was incorrect in \textsc{Type I} bugs. This brevity and clarity made it easier to evaluate their correctness and integrate their outputs into automated workflows.

\subsection{Threats to Validity}
\label{sec:threats}

Some threats to validity are important to be considered in this kind of work~\cite{sallou2024breaking}.
A potential threat to internal validity arises from the accuracy of our experimental setup, particularly in how we structured the prompts for evaluating SLMs. While we designed the prompts to be neutral and consistent, slight variations in wording could influence the models’ responses. Additionally, our evaluation relies on predefined ground truth classifications for refactoring bugs, meaning any mislabeling in this dataset could impact the reported detection rates. Another concern is the potential variability in model outputs due to non-deterministic behavior, particularly in API-based models like \omini{}, which may generate different responses when queried multiple times under similar preconditions.
The manual analysis of model responses presents an additional validity threat, as human bias and subjectivity could affect the interpretation of explanations and transformations. To mitigate this risk, we employed a structured evaluation process and cross-verified ambiguous cases to ensure consistency in assessments.

The generalizability of our findings is a key external validity concern. Our study focuses on Java and Python refactoring bugs, but results may not directly extend to other programming languages or refactoring types. While our approach is flexible and adaptable, different languages have unique syntax rules and constraints that may impact model performance. Additionally, our dataset is derived from bug reports in popular IDEs, yet it may not comprehensively cover all possible refactoring issues encountered in diverse development environments. Further validation across additional programming languages and larger datasets would enhance the robustness of our conclusions.

A threat to construct validity stems from our evaluation metrics and how we define successful bug detection. We assess models based on whether they correctly identify refactoring bugs, but this binary classification may not capture nuances such as partial correctness, uncertain responses, or explanations that are technically correct but vague. Furthermore, some models may correctly identify issues but fail to provide meaningful justifications, which we do not explicitly quantify in our analysis. Future work could refine evaluation criteria by incorporating confidence scores, explainability assessments, or expert validation of model responses.

Conclusion validity concerns whether our results are statistically and methodologically sound. Given that we analyze a limited set of refactoring bugs, there is a risk that our findings may be influenced by dataset biases rather than general trends in SLM performance. Additionally, differences in model architectures, training data, and reasoning capabilities could contribute to observed performance disparities. To strengthen our conclusions, we ensured a fair comparison by using the same prompts and evaluation criteria across all models. Further studies with expanded datasets and alternative model configurations could provide deeper insights into the reliability of SLMs for refactoring bug detection.

\section{Related Work}
\label{sec:relwork}

Opdyke and Johnson~\cite{Opdyke-SOOPPA-1990,Opdyke-PHD-1992} introduced the concept of refactoring, while Roberts~\cite{Roberts-PHD-1999} automated basic refactorings. Tokuda and Batory~\cite{Tokuda-ASE-2001} later showed that Opdyke’s preconditions alone were insufficient to guarantee behavior preservation. Sch\"{a}fer et al.~\cite{Schafer-PLPV-2009} further highlighted the challenges of guaranteeing correctness for all language constructs. 
Unlike these foundational works that focus on formalizing or proving refactorings, our approach employs SLMs to detect errors in existing tool implementations, thus offering an alternative path to ensuring correctness.

AlOmar et al.~\cite{DBLP:journals/infsof/AlOmarMNO21} performed a systematic mapping study on behavior preservation during software refactoring, providing a comprehensive overview of current practices, challenges, and research gaps in the field.
Drienyovszky et al.~\cite{quickcheck-erlang} employs an automated property-based random testing approach to validate Erlang refactoring tools by systematically checking behavior preservation properties.
Daniel et al.~\cite{test-tools-fse07} proposed an approach for automated testing of refactoring engines by detecting issues with overly weak preconditions using a Java program generator and a set of programmatic oracles. 
Soares et al.~\cite{Soares-TSE-2013} used \jdolly{} to generate Java programs and \saferefactor{}~\cite{saferefactor-ieee} to detect behavior changes. They found 106 bugs (compilation errors and behavioral changes) across 39 refactoring implementations. 
In contrast, our work examines the ability of SLMs to detect these same classes of bugs (\textsc{Type I}), thus extending earlier analyses with a new lightweight approach that does not rely on exhaustive program generation.

Daniel et al.~\cite{daniel-fse-07} and Jagannath et al.~\cite{Jagannath-FASE-2009} explored automated testing of refactoring implementations via program generators (\textsc{ASTGen} and STG) and various oracles, uncovering compilation errors or refactoring engine crashes. Steimann and Thies~\cite{steimann-ecoop} identified flawed refactoring implementations in mainstream Java IDEs, especially regarding accessibility. Gligoric et al.~\cite{Gligoric-ICSE-2010} introduced \textsc{UDITA} to make generator-based testing more expressive. 
While these approaches require manual setup for program generation, our method leverages SLMs, reducing the need for manual configuration while achieving similar goals of detecting compilation and behavioral errors.

Kim et al.~\cite{Kim-TSE-2014} investigated refactoring practices at Microsoft, revealing that developers often apply refactorings manually, except for the commonly automated Rename refactoring. Their findings underscore the need for more robust automated tools, aligning with our motivation to employ SLMs for more reliable refactoring support. Sch\"afer et al.~\cite{Schafer-OOPSLA-2010} improved \eclipse{} refactoring implementations by formalizing transformations, yet manual and language-specific effort remains a challenge. 
Conversely, our SLM-based strategy aims to minimize the need for manual intervention. Future improvements could involve refining our prompts by incorporating formalized properties for some refactoring types from their approach.

Bavota et al.~\cite{Bavota-scam-2012} conducted an empirical study linking refactoring activities to bug fixing, detecting 52 types of refactoring operations that induced fixes. Ketkar et al.~\cite{Ketkar-icse-12} presented a technique for learning and performing type changes in Java code, showcasing advanced automation capabilities. 
These works focus on specific refactoring categories or transformations, whereas our approach targets a broader range of refactorings, using SLMs to generalize across different languages and refactoring types.

Mongiovi et al.~\cite{Mongiovi-TSE-2018} and Soares et al.~\cite{Soares-ICSM-2011} proposed techniques to detect bugs caused by overly strong preconditions in refactoring engines (\textsc{Type II} bugs). One approach relies on generating small Java programs and modifying the refactoring engine code to disable certain preconditions. Similarly, Soares et al.~\cite{Soares-TSE-2013} identified several bugs in \eclipse{}’s refactoring implementations related to \textsc{Type I} bugs. 
Our work complements these generator-based techniques by introducing a lightweight SLM-based solution that scales to multiple languages (Java and Python) and does not require extensive code instrumentation.

Murphy-Hill et al.~\cite{Murphy-Hill-icse-2008} characterized refactoring problems in mainstream Java IDEs, revealing that poor error communication often leads to slow, conservative, or incorrect refactorings. They developed specialized tools to improve speed, accuracy, and usability, deriving recommendations for better refactoring support. 
Eilertsen and Murphy~\cite{refactoring-usability} introduce a theory -- based on an ISO definition of usability -- that aims to explain and improve the usability of refactoring tools. Through a study with 17 developers completing three refactoring tasks, the authors identified four themes that underscore developers' needs and preferences: more informative feedback from tools and greater control over how these tools operate. The paper refines its usability theory based on these findings, illustrating how a user-centered approach can uncover gaps between what tool designers assume (e.g., prioritizing safety) and what developers actually value (e.g., reduced invocation costs, even if less safe).
Our work complements existing efforts that focus on evaluating program transformations. Developers can adjust our natural-language prompts to adopt less constrained notions of behavior preservation. Furthermore, it would be valuable to extend the previous approach to evaluate the usability of AI-based IDEs that leverage foundation models, potentially providing new insights into how best to design and integrate automated refactoring capabilities.

Xu et al.~\cite{DBLP:conf/pldi/Xu0NH22} perform a systematic evaluation of LLMs for code. Hou et al.~\cite{se-llms-2023} presented a comprehensive systematic literature review on the application of LLMs in software engineering tasks. By examining 395 studies, their work classifies different models, data handling methods, and evaluation techniques, offering a broad overview of LLM usage in software engineering. Although their review includes multiple software engineering  challenges, it does not focus on refactoring bugs specifically. 
Fan et al.~\cite{DBLP:conf/fose-ws/FanGHLSYZ23} have highlighted the increasing use of LLMs in computer science, and some open problems in the refactoring area. 

Wei et al.~\cite{DBLP:conf/sigsoft/0003X023} have shown how LLMs can assist developers in program repair.
Ahmad et al.~\cite{DBLP:journals/tifs/AhmadTTKP24} show that LLMs have the ability to repair hardware security bugs.
Fan et al.~\cite{DBLP:conf/icse/FanGMRT23} investigate the extent to which Automatic Program Repair techniques can improve the reliability of code generated by LLMs. These results could complement our methodology by helping to correct some of the incorrect responses produced by SLMs in \textsc{Type II} bugs.

Salou et al.~\cite{sallou2024breaking} addressed the risks of using LLMs in SE, emphasizing threats like data leakage, reproducibility issues, and dependency on closed-source models. They propose guidelines to mitigate these risks and underscore the importance of empirical validation. 
Zhang et al.~\cite{zhang2023siren} provided an in-depth survey on hallucinations in LLMs, categorizing them into input-conflicting, context-conflicting, and fact-conflicting. Their taxonomy sheds light on potential pitfalls when generating or evaluating model outputs. 
We applied metamorphic testing~\cite{metamorphic-testing-2} to mitigate some of the threats.

Dong et al.~\cite{dong-icse-2025} propose a ChatGPT-based approach for testing refactoring engines. It uses ChatGPT to generate test programs aimed at uncovering defects. Their method builds a feature library extracted from existing bug reports and test cases, defines preconditions for each refactoring type, and uses predefined prompt templates to guide the automatic generation of test programs. These programs are subsequently employed in differential testing across multiple refactoring engines and manually analyzed to identify defects. Evaluating seven distinct refactoring types, the authors discovered a total of 115 \textsc{Type I} bugs in Java implementations.
In contrast, our approach differs from the previous work by focusing on analyzing the correctness (\textsc{Type I} and \textsc{Type II} bugs) of refactoring transformations rather than refactoring implementations. Our approach can detect behavioral changes, which are typically more challenging to identify. Additionally, our proposed method leverages automation more extensively and provides initial evidence supporting its applicability not only in Java, but also in Python and C, thereby indicating its broader versatility across different programming languages.

Dilhara et al.~\cite{DBLP:journals/pacmse/DilharaBBD24} propose PyCraft that integrates static and dynamic code analysis with LLM capabilities to refine code transformations for Python. PyCraft generates diverse code variations and corresponding test cases. To validate its effectiveness, the authors submitted 86 transformed code instances through 44 pull requests to projects with an acceptance rate of 83\%. These results highlight the potential of combining LLMs with analytical techniques for automated code transformation. 
Our approach can be employed to assess the correctness of the transformations proposed in their work, and detect \textsc{Type II} bugs.

Wang et al.~\cite{wang2024empiricalstudyrefactoringengine} conducted a manual analysis of 518 bugs across three popular refactoring engines, identifying key root causes, bug symptoms, and input program characteristics. Their study resulted in some findings, which were used to formulate a set of guidelines for refactoring engine bug detection and debugging. Additionally, their transferability study uncovered 130 new and unique bugs in the latest versions of refactoring tools, highlighting the pervasiveness of refactoring-related defects.
In contrast, our work evaluates a technique for identifying refactoring transformation bugs using SLMs. We assess our approach on \bugs{} reported bugs from widely used refactoring engines, some of which overlap with the dataset studied before. As future work, we aim to expand our evaluation by incorporating additional bugs from their dataset to further validate the effectiveness of our approach.

White et al.~\cite{white2023chatgptpromptpatternsimproving} propose two prompt templates (Pseudo-code Refactoring Pattern and The Data-guided Refactoring Pattern) for applying refactorings using ChatGPT.
AlOmar et al.~\cite{DBLP:conf/msr/AlOmarVMNO24} investigate how developers utilize ChatGPT for refactoring by analyzing developer-AI conversations. Through text mining techniques, the authors examine explicit refactoring intentions and the way ChatGPT responds to developers' needs. Their findings indicate that developers often make generic refactoring requests, while ChatGPT tends to infer and include specific refactoring intentions in its responses. 

Shirafuji et al.~\cite{DBLP:conf/apsec/ShirafujiOSMW23} propose a methodology for leveraging GPT-3.5 to suggest less complex versions of user-written Python programs through few-shot prompting with carefully selected examples. The quantitative evaluation indicates that the LLM successfully produces correct and simplified versions for 95.68\% of input programs, reducing cyclomatic complexity by 17.35\% and the number of lines of code by 25.84\%. Additionally, qualitative analysis reveals improvements in code readability through variable renaming and formatting but also highlights cases where the LLM introduces unnecessary modifications or removes comments. The findings demonstrate the potential of LLMs in code refactoring while also identifying their limitations, paving the way for future research on refining automated complexity reduction techniques.

Pomian et al.~\cite{DBLP:conf/sigsoft/PomianBDKBSBD24} propose an IntelliJ IDEA plugin called EM-Assist that generates, validates, enhances, and ranks Extract Method refactoring suggestions using LLMs. Unlike traditional static analysis-based refactoring tools, EM-Assist leverages AI to provide more context-aware recommendations. In an evaluation of 1,752 real-world refactorings from open-source projects, EM-Assist achieved a recall rate of 53.4\% among its top-5 recommendations. 
Depalma et al.~\cite{DBLP:journals/eswa/DepalmaMHMA24} present insights into the effectiveness and challenges of using ChatGPT for refactoring, highlighting both its strengths and limitations. Additionally, the study explores the ethical implications of AI-driven code transformations, addressing concerns related to biases, privacy, and potential risks in adopting these tools. Beyond evaluation, the work also offers practical recommendations for improving developer-AI collaboration in refactoring tasks. 
In contrast, our work focuses on assessing the correctness and reliability of AI-generated refactorings, emphasizing their impact on code behavior and transformation quality rather than on conversational patterns and ethical concerns.

Zhang et al.~\cite{ZHANG2025121753} apply deep learning and LLM-generated information to recommend Move Method refactorings. They constructs a dataset from 58 real-world projects, extracting metric features through static analysis, textual features using LLMs for code summarization, and semantic features by computing similarity between original and target classes. The dataset contains 12,475 samples and is used to train a CNN-LSTM-GRU model for refactoring recommendation. Experimental results show that their tool achieves an average F1-score of 74\%, outperforming existing tools such as PathMove, JDeodorant, JMove, and RMove, with improvements ranging from 9.4\% to 53.4\%. These results highlight the potential of combining deep learning with LLM-generated insights to enhance refactoring automation. 
Liu et al.~\cite{DBLP:conf/issta/LiuWWXWLJ23} introduce RefBERT, a two-stage pre-trained framework for automatic variable renaming based on the BERT architecture. In the first stage, RefBERT predicts the number of sub-tokens required for the new name; in the second, it generates the sub-tokens accordingly. The framework integrates several advanced techniques to effectively address the unique challenges of this refactoring task. Experimental evaluations on constructed refactoring datasets reveal that RefBERT surpasses existing methods, producing variable names that are both accurate and meaningful.
Our work focuses on ensuring the correctness of any type of refactorings using SLMs.

Cordeiro et al.~\cite{cordeiro2024empiricalstudycoderefactoring} present an empirical study comparing \emph{StarCoder2} -- an LLM tailored for code generation -- to human developers in code refactoring tasks. The authors analyze 30 open-source Java projects, evaluating the ability of StarCoder2 and developers to reduce various types of code smells. Their results show that StarCoder2 achieves a higher overall reduction rate of code smells than developers, particularly excelling at systematic, repetitive refactorings. In contrast, human developers outperform the model on complex, context-dependent design issues.
Our work complements the above by focusing on ensuring the correctness of program transformations using SLMs.

\section{Conclusions}
\label{sec:conclusion}

In this work, we investigated the effectiveness of Small Language Models (SLMs) in detecting refactoring bugs in Java and Python, including both incorrect transformations (\textsc{Type I}) and valid refactorings that are incorrectly rejected (\textsc{Type II}). Using a curated dataset of real-world bugs from popular IDEs, our results show that SLMs can accurately identify a substantial portion of these issues. Notably, the proprietary \ominihigh{} model achieved the highest detection rate, while the open-source \phimodel{} delivered comparable performance, demonstrating the promise of accessible, lightweight models in supporting reliable refactoring automation.

Our empirical results underscore the capability of SLMs to detect refactoring bugs -- particularly \textsc{Type I} -- with high accuracy. The \phimodel{} correctly identified 71.4\% of these bugs, and the \ominihigh{} model reached 84.3\%. Although the detection of \textsc{Type II} bugs was more limited (with a maximum of 40\%), the approach requires neither formal preconditions nor complex instrumentation, and generalizes well across programming languages (Java, Python, and C) and refactoring types. \phimodel{} also demonstrated robustness to structural variations in the code, as shown in our metamorphic testing. Furthermore, we observed high \textit{pass@\(k\)} values with few attempts and an average consistency of 78.07\%, reinforcing the practicality of \phimodel{} as a scalable tool for validating refactoring correctness. Future work could benefit from a mixture-of-experts strategy, combining complementary model strengths to further improve accuracy and coverage.

An important aspect of our study was the manual evaluation of all model-generated explanations. Without this verification step, the detection rates reported in Section~\ref{sec:results} would be artificially inflated. Human oversight remains essential for validating model reasoning, ensuring that explanations are trustworthy and that software engineers are not misled. This highlights the importance of keeping developers in the loop to verify and refine model outputs -- an essential component of reliable, human-in-the-loop AI-assisted software maintenance.

SLMs equipped with type checkers proved highly effective in identifying most \textsc{Type I} bugs. However, detecting \textsc{Type II} bugs remains a significant challenge, with current models achieving only moderate success. In contrast, traditional techniques~\cite{Mongiovi-TSE-2018,Soares-ICSM-2011} -- though more labor and resource-intensive --achieve full detection by analyzing and relaxing overly strict preconditions embedded in refactoring tools. Our dataset thus serves as a benchmark for tracking progress in addressing such complex scenarios using language models.

Beyond the refactoring domain, our findings have broader implications for other areas of software engineering. Compiler optimizations, which rely on behavior-preserving transformations, could benefit from the same principles. Similarly, mutation testing~\cite{demillo1978hints,pizzoleto-systematic-mutation-jss19,DBLP:conf/icse/SteimannT10,rohit-subsumption-ist21,marcio-subsumption-icst-2020} may leverage SLMs to identify equivalent and redundant mutants more efficiently, reducing the cost of non-informative test executions. By minimizing dependence on static or dynamic analysis, SLMs offer the potential for faster feedback loops and lower operational costs, making them attractive for both academic and industrial settings.

As future directions, we plan to expand the dataset and include additional programming languages to evaluate the generalizability of our approach. We aim to assess SLM performance on larger codebases and explore applications in compiler optimization and mutation testing, leveraging similar principles of efficient bug detection and automated validation. We also intend to investigate different model configurations, fine-tune models like \deepseek{} for refactoring-specific tasks, and explore agentic strategies that integrate external tools to enhance reasoning and detection accuracy. Finally, we aim to develop a distilled, specialized version of a selected model fine-tuned for refactoring-related bug detection, enabling broader adoption and more precise analysis across a variety of software engineering scenarios.

\section*{Acknowledgments}
This work was partially supported by CNPq, FAPESQ-PB and FAPEAL grants.

\end{document}